# Tri-Functional Metasurface for Phase, Amplitude, and Luminescence Control

Soroosh Daqiqeh Rezaei[1,2,6,†,*], Zhaogang Dong[3,†], Hao Wang[1], Jiahui Xu[4], Hongtao Wang[1], Mohammad Tavakkoli Yaraki[5], Ken Choon Hwa Goh[3], Wang Zhang[1], Xiaogang Liu[3,4], and Joel K. W. Yang[1,3,*]

[1] Pillar of Engineering Product Development, Singapore University of Technology and Design (SUTD), 8 Somapah Road, 487372, Singapore

[2] Connection One, METATAG PTE. LTD., 150167, Singapore

[3] Institute of Materials Research and Engineering (IMRE), A*STAR (Agency for Science, Technology and Research), 2 Fusionopolis Way, 138634, Singapore

[4] Department of Chemistry, National University of Singapore, Singapore 117543, Singapore

[5] School of Natural Sciences, Macquarie University, Sydney, NSW 2109, Australia

[6] Present address: Department of Electrical Engineering, The Pennsylvania State University, University Park, PA, 16802, USA

E-mail: soroosh@u.nus.edu, joel_yang@sutd.edu.sg

† Contributed equally

Keywords: Structural color, hologram, luminescence enhancement, upconversion enhancement, optical anti-counterfeiting

In optical anti-counterfeiting, several distinct optically variable devices (OVDs) are often concurrently employed to compensate for the insufficient security level of constituent OVDs. Alternatively, metasurfaces that exhibit multiple optical responses effectively combine multiple OVDs into one, thus significantly enhancing their security and hindering fraudulent replication. This work demonstrates the simultaneous control of three separate optical responses, *i.e.,* phase, amplitude, and luminescence, using anisotropic gap-plasmon metasurfaces. Due to the incorporated geometric anisotropy, the designed structure exhibits distinct responses under *x*- and *y*-polarized light, revealing either a color image, or a holographic projection in the far field. Furthermore, inserting upconversion nanoparticles (UCNPs) into the dielectric gaps of the structures, the designed metasurface is able to generate a third luminescent image upon illumination with the near-infrared light. The stochastic distribution of the UCNPs constitutes a unique "fingerprint", achieving a physically unclonable function (PUF) layer. Crucially, our triple-mode metasurface requires only readily attainable equipment such as a macro-lens/camera and a laser pointer to read most of the channels, thus paving the way towards highly secure and easy-to-authenticate metasurface-driven OVDs (mOVDs).

## 1. Introduction

Based on the unique interactions of light with nanostructures, optical metasurfaces underpin an extensive array of applications ranging from neuromorphic computing[1–5] to holography,[6–11] orbital angular momentum generation (OAM),[12–18] structural color printing,[19–25] and anti-counterfeiting.[26–32] Among the plethora of use cases, metasurfaces that enable manipulation of more than two types of optical properties are paramount to creating metasurface-driven optically variable devices (mOVDs) via combining several security features into one, placing a damper on the rise of counterfeit goods.[33]

The two most common OVDs used for anti-counterfeiting are color prints and holograms.[34] Several works report on integrating these two features by simultaneous control of phase and amplitude to generate holographic and color images using metasurfaces. Such dual-mode operation is often achieved by employing Mie resonances in high refractive-index materials in combination with the geometric phase.[35,36] In this approach, the size of building blocks determines the spectral response while their orientation can tailor the designed phase distribution using a circularly polarized source.[37,38] Yet in other works, 3D printed phase plates, and waveguide modes in low-refractive-index nanopillars have been employed successfully to modulate phase and amplitude.[39,40] Another common approach integrates Fabry-Perot color filters with other resonators to control amplitude and phase.[41–43]

Alternatively, plasmonic resonators can be employed to manipulate phase and amplitude to realize mOVDs.[44] Particularly gap-plasmon resonators (GPRs) benefit from strong field confinements,[45] thus achieving ultra-high-resolutions[46] not attainable by other systems. So far, GPRs have been employed to tailor various light properties such as phase,[47–49] amplitude,[50,51] luminescence,[52] polarization,[53–56] and been utilized in applications like beam steering.[57–60] Therefore, GPRs are a prime candidate to merge multiple functionalities and strengthen the security of OVDs. An ideal OVD should (1) leverage on multiple optical effects, (2) be easy to read for most modes, (3) work in reflection as most tagged assets are not transparent, (4) possess a high field of view, (5) be made from robust materials, and (6) be planar and compact.

Thus far, dual-mode mOVDs for concurrent control of phase-amplitude, or amplitude-luminescence have been reported. However, simultaneous control of phase-amplitude-luminescence remains elusive. Here, we employ anisotropic gap-plasmon structures with integrated upconversion nanoparticles (UCNPs) to demonstrate simultaneous control of phase, amplitude and luminescence, creating a triple-mode mOVD with overt, covert, and forensic features. The overt security feature is displayed under the white light source as a microprint, while the covert and forensic features are revealed under red and near-infrared lasers as a hologram and luminescence image. Moreover, the stochastic distribution of UCNPs in the metasurface creates a PUF feature, rendering the mOVD physically unclonable. Not only combining multiple features drastically enhance the complexity for replication, it also improves concealment as a single tag is used instead of multiple tags, further preventing forgery. While most of the reported works so far offer concurrent phase and amplitude control, in practice these systems are not in practice suited for easy readout due to the complicated optical setup required, *e.g.*, for circularly polarized source and analyzer. Crucially, from an application standpoint, our method enables ease of authentication using commonly available tools, *e.g.*, macro-lenses and laser pointers, while preserving covertness. In addition, our developed mOVD is fabricated from robust materials, is planar (~ 170 nm thick), and shows good angle insensitivity for the color print and the hologram, further expediting adoption.

## 2. Results & Discussion

We fabricated arrays of nanostructures, one of which is schematically shown in **Figure 1A.** Structures were fabricated by thin film deposition and patterning using electron beam lithography (EBL) as described in the Experimental section. Our proposed design relies on the anisotropy of structures with respect to *x* and *y* polarizations. The cross-shaped gap-plasmon structure consists of a 30-nm-thick $SiO_2$ film sandwiched between the top 40-nm-thick Al layer and 100-nm-thick Al backreflector with an arm length of $L_{xx}$ = 170 nm and $L_{yy}$ = 126 nm. The scanning electron microscope (SEM) image in Figure 1B further depicts the anisotropy of the geometry. The dimensions of the structure fabricated in this work support strong gap-plasmon resonances, similar to previous reports in the literature.[61,62]

To independently control and tune the color by switching incident polarization, we fabricated the color palette in Figure 1C with varying $L_{yy}$ = 85 to 175 nm in steps of 5 nm for two armlengths $L_{xx}$ = 170 nm and $L_{xx}$ = 126 nm at constant pitch $P$ = 250 nm. $L_{xy}$ and $L_{yx}$ were set to 80 and 40 nm respectively to minimize the cross-talk between orthogonal polarizations while maintaining saturated colors and relatively high amplitude for a brighter hologram. The choice of geometrical parameters will be discussed in greater detail later. As can be observed under $x$-polarized light, with varying $L_{yy}$, top row remains cyan while the bottom row maintains the magenta color with small color variation in every row. The fixed color in each row albeit variation in $L_{yy}$ is due to the constant geometrical size along $x$-polarization. Next, the incident light polarization is switched to $y$ for the same color palette in Figure 1C. In this scenario the color transitions from yellow to cyan by increasing the armlength $L_{yy}$, redshifting the gap-plasmon resonance with increased length along the $y$-axis (Figure 2D). In this case the colors generated for any given $L_{yy}$ look strikingly identical for $L_{xx}$ = 126 and 170 nm, highlighting independent color control with polarization. This outcome showcases that the gap-plasmon resonance depends on the armlength along the polarized light rather than the arm orthogonal to the incident light polarization. In other words, the dimension along polarization dictates the resonance and the resulting color. To further investigate the optical properties of the color palette, reflectance measurements were carried out under orthogonal polarizations.

By measuring the reflectance spectra of the color palette in Figure 1C under $x$-polarized light as plotted in Figure 1E, it can be shown that in fact the resonance dips do not change for $L_{xx}$ = 170 nm and $L_{xx}$ = 126 nm while $L_{yy}$ is varied. The presence of dominant absorption dips in the reflectance spectra indicates the presence of the gap-plasmon resonance.[61] To confirm the existence of this resonance and shed light on the underlying physics inducing the absorption, the optical field distributions in the gap region are simulated at reflectance dips and plotted in orthogonal planes of the resonator.

Figure 1F depicts the Poynting fields at the reflectance dip (highlighted with the inverse cyan triangle in Figure 1C and E) of the color patch with $L_{xx}$ = 170 nm and $L_{yy}$ = 125 nm. As observed in the polarization plane, i.e., top $x$-$z$, light energy is concentrated in the long arm gap with the form of a standing wave and absorbed by Al, giving rise to the reflectance dips highlighted in Figure 1E. The confinement of light energy and the formation of a standing wave in the gap (see **Figure S1** for electric-field simulation) confirm the presence of the gap-plasmon mode.[63] Conversely, the light energy concentration in the structure gap for the orthogonal plane $x$-$y$ in Figure 1F is negligible. This low energy concentration indicates that no resonance is present in the plane of the shorter arm. However, when the polarization is switched to $y$, the gap-plasmon resonance is activated in the $y$-$z$ plane and tuned by varying $L_{yy}$ as elucidated in the reflectance spectra in Figure 1G. Similar to the previous field simulation, the resonance can be observed only in the plane of polarization, *i.e.*, Figure 1H *y-z*, at the highlighted reflectance dip while the orthogonal polarization plane Figure 1H *x-z* shows no sign of a resonance. Therefore, by switching the incident polarization, gap-plasmon resonances can be switched on and off independently, giving rise to absorption in the visible. This effect is later employed to switch and tune the reflected amplitude and phase in orthogonal planes for color generation and holography.

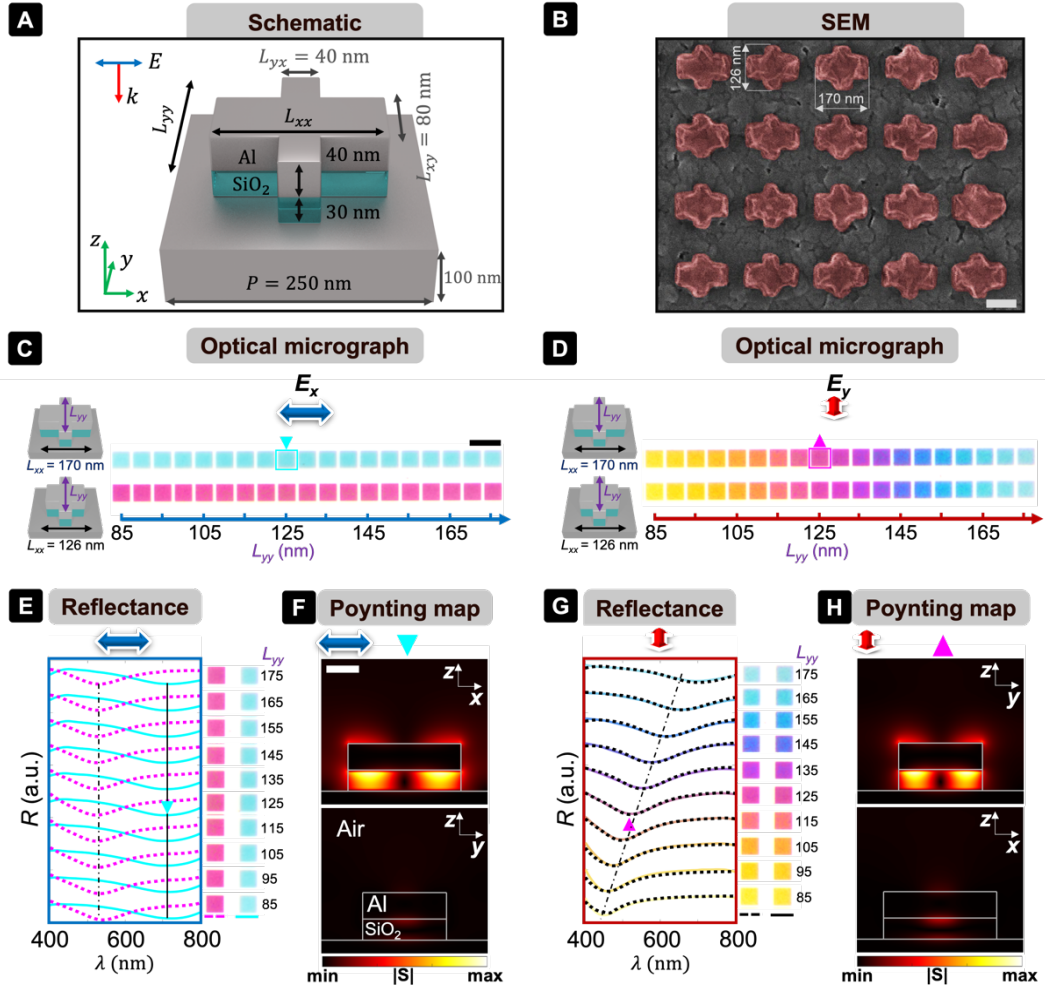

**Figure 1.** (**A**) Schematic of cross-shaped nanostructure supporting gap-plasmons. (**B**) False colorized scanning electron microscope image of the metasurface fabricated by electron beam lithography. (**C**) Optical micrographs of the color palette of arrays with $L_{xx}$ = 126 nm and $L_{xx}$ = 170 nm for $L_{yy}$ = 85:5:175 nm under *x*-polarized and (**D**) *y*-polarized incidence. The pitch is set to $P$ = 250 nm for all arrays. (**E**) Reflectance spectra of selected arrays with $L_{xx}$ = 126 nm and $L_{xx}$ = 170 nm under *x*-polarized light. (**F**) Poynting field distribution of the selected array in the *x-z* plane (top) and *y-z* plane (bottom) under *x*-polarized light. (**G**) Reflectance spectra of selected arrays with $L_{xx}$ = 126 nm and $L_{xx}$ = 170 nm under *y*-polarized light. (**H**) Poynting field

distribution of the selected array in the *y-z* plane (top) and *x-z* plane (bottom) under *y*-polarized light. Scale bars in (**B**), (**C**), and (**F**) are 100 nm, 20 μm, and 50 nm respectively.

To achieve the independent phase and amplitude control in orthogonal linear polarization planes, we simulated the effect of armlength variation on reflected phase as plotted in **Figure 2A**. Similar to the color palette in Figure 1C the array periodicity is set to $P = 250$ nm with $L_{yy} = 170$ nm, $L_{xy} = 80$ nm, and $L_{yx} = 40$ nm. It can be seen from the phase map that for a given wavelength, a difference of ~45 nm in armlength can yield a $\pi$-phase difference. This phase difference is the first step to design a binary hologram with linear polarization and constant amplitude. To find the optimum unit cell design for both color printing and holography we studied various geometrical parameters (**Figure S2-5**). In fact, the design used throughout this paper is based on this optimization. For the red laser with $\lambda = 638$ nm, armlengths $L_{xx} = 126$ nm and $L_{xx} = 170$ nm yield the required $\pi$-phase separation with equal reflected amplitude according to plots in Figure 2A and B. Similar to the independent color control shown earlier in Figure 1C and Figure 1D, a hologram can be encoded by setting $L_{xx}$ to 126 nm and 170 nm to create the phase pixels and varying $L_{yy}$ to generate different colors. With such approach a hologram can be produced under *x*-polarized monochromatic light while *y*-polarization can generate a color image, creating a dual-mode metasurface.

The dual-mode metasurface is shown in Figure 2C. Under *y*-polarized white light a color print is revealed. By illuminating the metasurface with an *x*-polarized red laser, a hologram is projected into the far-field as captured in Figure 2D using the optical setup in **Figure S6**. The SEM micrograph of the metasurface is displayed in Figure 2E highlighting the anisotropy of the design. This approach enables storing an independent hologram and color image in orthogonal linear polarizations. The results agree with the simulated optical micrograph and hologram in **Figure S7**. Next, we attempt to utilize an additional mode, *i.e.*, illuminance, by incorporating UCNPs into the dielectric gap of our design.

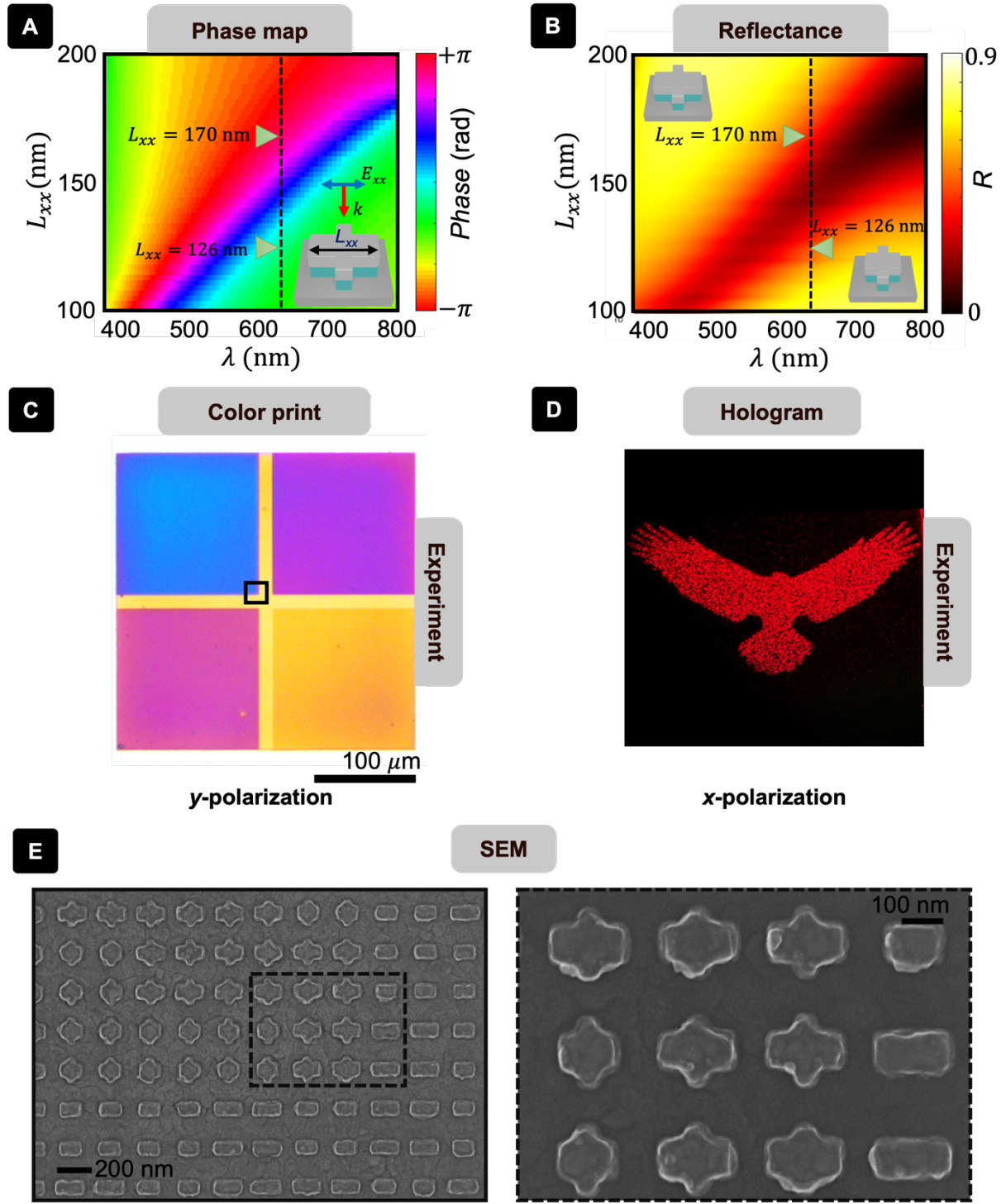

**Figure 2.** (**A**) Simulated reflectance phase map distribution for various armlength $L_{xx}$ in the visible spectra under *x*-polarized incident light. With $P = 250$ nm, $L_{yy} = 170$ nm, $L_{xy} = 80$ nm, $L_{yx} = 40$ nm. The laser wavelength $\lambda = 638$ nm is highlighted with a dashed line. (**B**) Simulated reflectance map versus the armlength $L_{xx}$. (**C**) Optical micrograph of the bi-functional metasurface under *y*-polarized white light. (**D**) Experimental holographic image captured with a camera in the far-field by illuminating the bi-functional metasurface sample with an *x*-

polarized red laser at $\lambda = 638$ nm wavelength. (**E**) SEM image of the bi-functional metasurface sample with a higher magnification inset.

The UCNPs utilized here, absorb the 980 nm near-infrared pump laser and emit through two main emission peaks in the visible spectra located at 545 and 655 nm.[52] To incorporate UCNPs into the dielectric gap, after EBL and resist development, UCNPs were spin-coated followed by electron beam evaporation (EBE) as detailed in Experimental section. Such array is depicted under $x$ and $y$-polarized illumination in **Figure 3A-B**. Similar to the array in Figure 1C, the colors can be independently controlled by switching the polarization. Notably, the colors are only slightly redshifted compared to the case without UCNPs. The colors can also be observed under unpolarized source as illustrated in Figure 3C. By illuminating the sample under unpolarized near-infrared excitation at 980 nm, a luminescence image of the array can be observed in Figure 3D-F, demonstrating the presence of upconverted emission. Luminescence array colors ranges from green to blue that matches with the emission peaks of the UCNPs at 545 and 655 nm. The UCNPs are incorporated into the gap of the resonators including the region of enhanced electromagnetic fields of the gap-plasmon mode. As a result, the UCNPs experience a large Purcell factor enhancement through plasmon-emitter coupling.[52] By tuning the size of the gap-plasmon resonators, different emission channels can be enhanced, resulting in the luminescence color tuning observed in Figure 3F.

When the near-infrared source is $y$-polarized, the patches gradually transition from dark to bright with increasing $L_{yy}$ (Figure 3D). The increase in brightness is consistent with the results reported in previous work.[52] with the increase in $L_{yy}$, the absorption dip of the gap-plasmon resonator redshifts, approaching the pump wavelength at 980 nm. With more photons absorbed, the upconversion process is enhanced, resulting in a brighter luminescence image. It is worth noting that both rows in Figure 3D appear almost equally bright, as the $L_{yy}$ is the same for both. The slightly darker patch in the center is caused by the unevenness of the laser illumination. By switching the polarization to $x$, the upper row appears brighter as the resonance dip for $L_{xx} = 170$ nm is closer to the pump wavelength (Figure 3E). Apart from the uneven illumination, the brightness is almost constant for every row as $L_{xx}$ is constant for each row. Consequently, Figure 3D and Figure 3E highlight that luminescence can be controlled by polarization as well. When the array is illuminated with an unpolarized source, the brightness is considerably improved for all patches as there is no polarizer to absorb the energy (Figure 3F), hence the upper row appears brighter. This observation can be explained if we assume that the unpolarized source is similar to the superposition of $x$ and $y$-polarized sources. As the row with $L_{xx} = 170$ nm was brighter for linear polarizations, it should also be brighter for the unpolarized case. Figure 3G left shows the assembled UCNPs inside the PMMA pattern. As can be seen, the assembly almost covers the whole Al backreflector. The enhanced filling factor is achieved by further optimizing the spin coating process reported in our previous work.[52] We have demonstrated independent color and phase control by switching polarization in addition to up-conversion illuminance. Based on results above, we next show how these modes can be employed in a single metasurface.

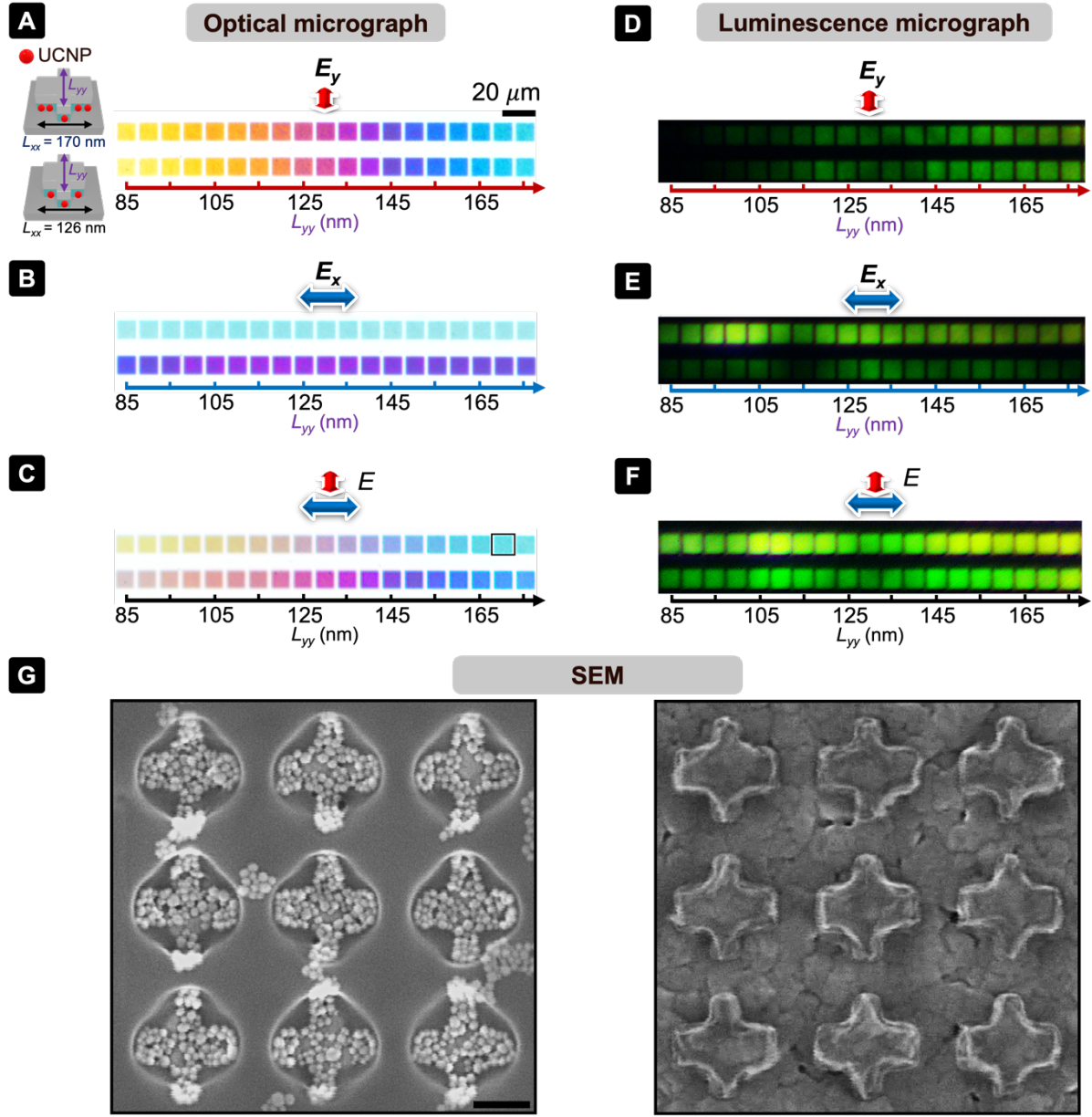

**Figure 3.** (**A**) Color palette of UCNPs incorporated arrays with $L_{xx}$ = 126 nm and $L_{xx}$ = 170 nm for $L_{yy}$ = 85:5:175 nm under $y$-polarized, (**B**) $x$-polarized illumination, and (**C**) unpolarized white light (**D**) luminescence image of the same arrays under $y$-polarized, (**E**) $x$-polarized, and (**F**) unpolarized near-infrared excitation at 980 nm. (**G**) SEM micrograph of the assembled

UCNPs in PMMA nanocrosses (left) and SEM of the same array after deposition of $SiO_2$ and Al capping layers. With $P$ = 250 nm, $L_{xx}$ = 170 nm, $L_{yy}$ = 170 nm, $L_{xy}$ = 80 nm, $L_{yx}$ = 40 nm. The scale bar in (**G**) is 100 nm.

The metasurface in **Figure 4A** is fabricated by incorporating UCNPs into the gap. As can be observed, a color image appears under *y*-polarized white light, while by pumping a near-infrared laser and illuminating with an *x*-polarized red laser produces a luminescence image and a far-field hologram, respectively (Figure 4B and Figure 4C). The experimental results are in good agreement with simulations in **Figure S8**. It is worthwhile highlighting that various luminescence images can be generated under various polarizations (**Figure S9**) which is consistent with results in Figure 3. Figure 4D shows the SEM image of the UCNPs in the sample before deposition of capping layers. The distribution nearly covers the exposed surface of the pattern, although it is not entirely uniform. This stochastic distribution of the UCNPs generates a unique "fingerprint", achieving a PUF layer. While the optical images of samples with nominally identical EBL exposures and process conditions are nearly identical as elucidated in Figure 4E top, due to the stochastic distribution of UCNPs, no two luminescence images can be identical (Figure 4E bottom), creating a PUF that further enhances security.

In addition to multiple functions and security levels, easy authentication is crucial for end-user in practical applications. The color image and hologram can be easily revealed by a macrolens and laser pointer without the need for a polarizer as demonstrated in **Figure S10,** providing the overt and convert security features. While the luminescence adds a forensic security feature. Finally, our design facilitates adoption for many use case scenarios as it works in reflection mode, which is more desirable than transmission mode as many objects that need to be tagged are not transparent. It is worthwhile mentioning that various strategies can be employed to extend this work to achieve multiple color holograms as reported in the literature before.[64]

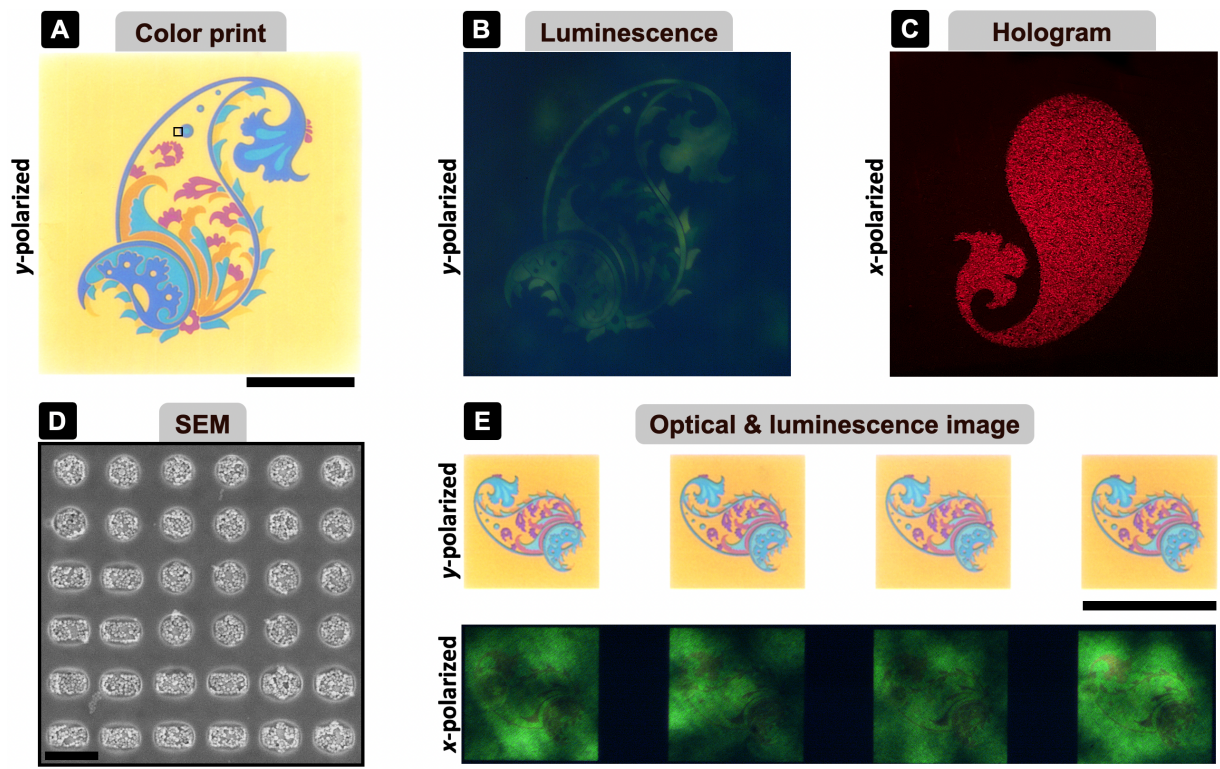

**Figure 4.** (**A**) Optical image of the tri-functional metasurface under *y*-polarized white source. (**B**) Luminescence image of the same metasurface under unpolarized near-infrared excitation of 980 nm. (**C**) The reflected holographic image captured in the far-field when the sample is illuminated with an *x*-polarized red laser beam at 638 nm. (**D**) SEM image of the area with UCNPs highlighted in panel (**A**) before deposition of $SiO_2$ and Al capping layers. (**E**) Optical and luminescence image of samples with nominally identical EBL exposure and process conditions under *y*-polarized white source (top) and *x*-polarized near-infrared excitation of 980 nm (bottom).

## 3. Conclusion

In summary, we demonstrated simultaneous control of phase, amplitude, and luminescence by employing anisotropic gap-plasmon structures with upconversion nanoparticles incorporated in their dielectric gap. The tri-functional metasurface generated a color image under white light while producing a luminescence image under a near-infrared source and a hologram in the far-field by red laser illumination. Moreover, the stochastic distribution of the UCNPs in the metasurface creates a PUF feature. Therefore, the design presented here offers a nearly unclonable security tag by combining overt, covert, and forensic security features in addition to the PUF. Prior works heavily rely on complicated optical setups and various linear and nonlinear polarizations, while our approach overcomes these complications and offers simplicity in reading different channels through a macro-lens and a laser pointer. Our work paves the way for creating unclonable tags for easy authentication.

## 4. Methods

*Preparation of UCNPs*: Ytterbium (III) acetate hydrate (99.9%), yttrium (III) acetate hydrate (99.9%), erbium (III) acetate hydrate (99.9%), oleic acid (90%), 1-octadecene (90%), sodium hydroxide (NaOH; >98%), ammonium fluoride ($NH_4F$; >98%) and cyclohexane were purchased from Sigma-Aldrich and used as received without further purification. $NaYF_4$:Yb/Er (18/2 mol%) nanocrystals were synthesized according to previous reports.[65] A 4-mL aqueous solution of Ln $(CH_3CO_2)_3$ (0.2 M, Ln= Y, Yb, and Er) was added to a 50-mL flask containing 6 mL of oleic acid and 14 mL of 1-octadecene. The reaction mixture was heated to 150 °C and kept for one h under stirring to remove water from the solution. After cooling to room temperature, a 12-mL methanol solution containing $NH_4F$ (3.2 mmol) and NaOH (2 mmol) was added to the mixture and kept under stirring for 30 min. After removing the methanol, the solution was heated at 290 °C under argon for 1.5 h and then cooled to room temperature. The resulting nanoparticles were washed with ethanol several times and re-dispersed in 4 mL of cyclohexane.

*Fabrication of Gap-Plasmon Structures*: First, a 100-nm-thick Al film was deposited on Si wafer using electron beam evaporation (Kurt J. Lesker, deposition rate of 1 Å s$^{-1}$). The thickness and evaporation rates were measured *in situ* with a quartz balance. Second, poly-methyl methacrylate (PMMA A5, diluted in anisole solvent with ratio of 1:1) resist (950k molecular weight, 3.3% wt in anisole) was spin coated onto an Al-coated Si substrate at a spin speed of 3k revolutions-per-minute (rpm) to obtain a ~100 nm thick PMMA layer. The PMMA resist was then baked at 180 °C for 120 s to remove the solvent. The sample was exposed by EBL (Elionix ELS-7000, electron acceleration voltage of 100 keV, beam current of 100 pA). Afterwards, the sample was developed in a solution of methyl isobutyl ketone (MIBK) and isopropyl alcohol (IPA) solution with a volume mixing ratio of 1:3, at a temperature of −10 °C for 15 s, followed by blow drying the sample with nitrogen.[66] Finally, 30-nm-thick $SiO_2$ and 40-nm-thick Al were deposited (with deposition rate of 1 Å s$^{-1}$), followed by a lift-off process (acetone at room temperature).

*Fabrication of Gap-Plasmon Structures with UCNPs*: First, a 100-nm-thick Al film was deposited on Si wafer using electron beam evaporation (Kurt J. Lesker, deposition rate of 1 Å s$^{-1}$). The thickness and evaporation rates were measured *in situ* with a quartz balance. Second, poly- methyl methacrylate (PMMA A5, diluted in anisole solvent with ratio of 1:1) resist (950k molecular weight, 3.3% wt in anisole) was spin coated onto an Al-coated Si substrate at a spin speed of 3k revolutions-per-minute (rpm) to obtain a ~100 nm thick PMMA layer. The PMMA resist was then baked at 180 °C for 120 s to remove the solvent. The sample was exposed by EBL (Elionix ELS-7000, electron acceleration voltage of 100 keV, beam current of 100 pA). Afterwards, the sample was developed in a solution of methyl isobutyl ketone (MIBK) and isopropyl alcohol (IPA) solution with a volume mixing ratio of 1:3, at a temperature of −10 °C for 15 s, followed by blow drying the sample with nitrogen. Next, the UCNPs water solution was spin coated onto the sample at a spin speed of 5k revolutions-per-minute (rpm) and baked at 90 °C for 1 min. In the end, 7.5-nm-thick $SiO_2$ and 40-nm-thick Al were deposited (with deposition rate of 1 Å $s^{-1}$), followed by a lift-off process (acetone at room temperature).

*SEM Characterization*: The SEM images were obtained using the eLINE Plus system (Raith) at an acceleration voltage of 10 kV and an aperture size of 30 μm.

*Optical Characterization*: Reflectance spectra were measured with a micro-spectrophotometer (CRAIC) equipped with a 20 × lens (NA = 0.45). The reflectivity of a 100-nm-thick Al mirror was measured as the reference signal. All the optical microscopy images were collected with the same microscope system.

*Upconversion Luminescence Imaging*: Upconversion luminescence microscopy images were collected on an Olympus BX51 microscope with a Xenon lamp adapted to a 980 nm CW laser and equipped with a 20 × lens (NA = 0.45). The excitation power was set to 60 mW. and images were scanned with a 1 s integration time.

*Hologram Design*: The binary hologram was designed by computer-generated hologram with an adaptive Fourier transform based algorithm. The Gerchberg-Saxton algorithm was adopted to get the final binary phase.[67] First, a plane wave (laser) passed through a random phase and a projection was obtained in the far-field; Second, the amplitude of the projection was replaced by the designed image and then inverse Fourier transform was performed; Third, in the hologram plane, the amplitude was replaced by the plane wave and then Fourier transform was carried out again. The signal to noise ratio stabilized after iterations and the phase was limited to be binary to fit in our design.

*Photography of Holographic Projections*: Holograms were projected in reflection onto a white screen and photographed using a DSLR camera in a darkened room. Coherent illumination was provided by 638 nm red laser diode module with a maximum power of 4.5 mW (ThorLabs). The distance of the holographic colour prints from the screen (projection distance) was 30 cm, at which the holographic projections measured between 10 and 15 cm across.

*Numerical Simulations*: Finite-difference time-domain simulations were carried out using a commercial software (Lumerical FDTD Solutions). To calculate the electric field and power loss map, a standard plane wave was employed with periodic boundary conditions along the *x*- and *y*-axes. Due to structural symmetry, symmetry conditions were used to reduce the computational time. Perfectly matched layers were used along the propagation direction at the top and bottom boundaries. The permittivity data of Al and $SiO_2$ were obtained from the literature.[68]

**Supporting Information**

Supporting Information is available from the Wiley Online Library or from the author.

**Acknowledgements**

This research is supported by National Research Foundation (NRF) Singapore, under its Competitive Research Programme NRF-CRP001-021 and CRP20-2017-0004, as well as NRF Investigatorship Award NRF-NRFI06-2020-0005. In addition, Z.D. would like to acknowledge the funding support from A*STAR AME IRG grant (Project No. A20E5c0093), A*STAR CDA grant (Project No. C210112019) and A*STAR MTC IRG grant (Project No. M21K2c0116).

**TOC**

This work demonstrates the simultaneous control of three separate optical responses, *i.e.,* phase, amplitude, and luminescence, using anisotropic gap-plasmon metasurfaces. Due to the incorporated geometric anisotropy, the designed structure exhibits distinct responses under *x*- and *y*-polarized light, revealing either a color image, or a holographic projection in the far field.

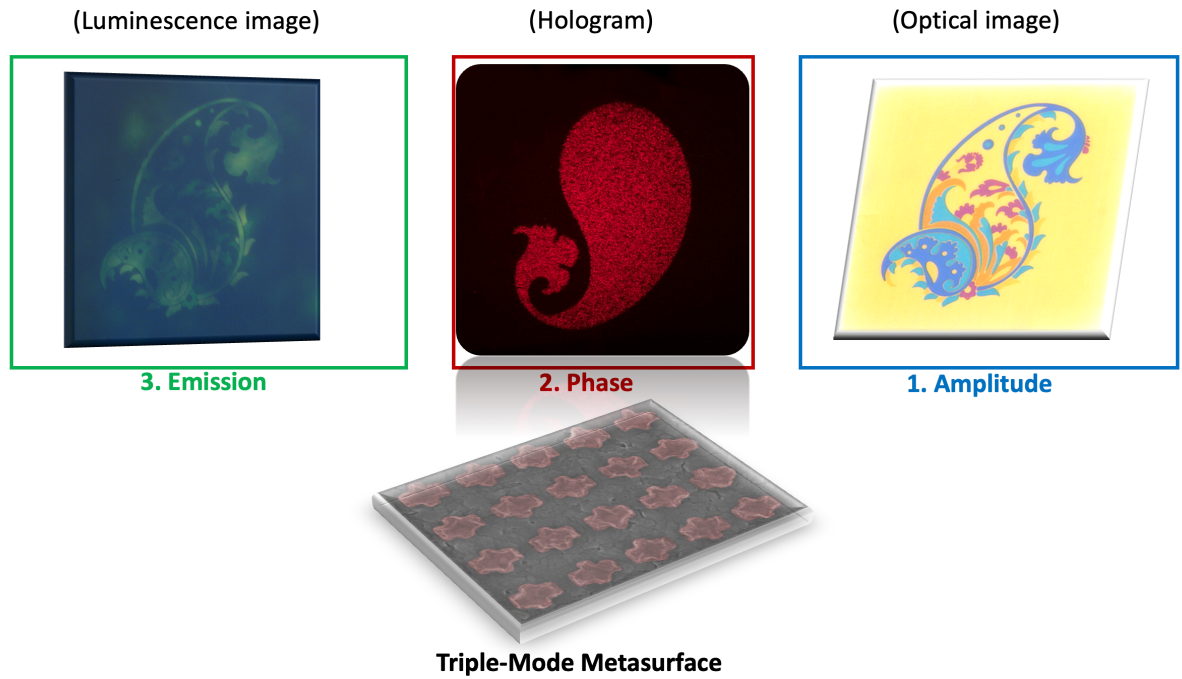

# Supporting Information

**Tri-Functional Metasurfaces for Phase, Amplitude, and Luminescence Control**

Soroosh Daqiqeh Rezaei[1,2,6,†,*], Zhaogang Dong[3,†], Hao Wang[1], Jiahui Xu[4], Hongtao Wang[1], Mohammad Tavakkoli Yaraki[5], Ken Choon Hwa Goh[3], Wang Zhang[1], Xiaogang Liu[3,4], and Joel K. W. Yang[1,3,*]

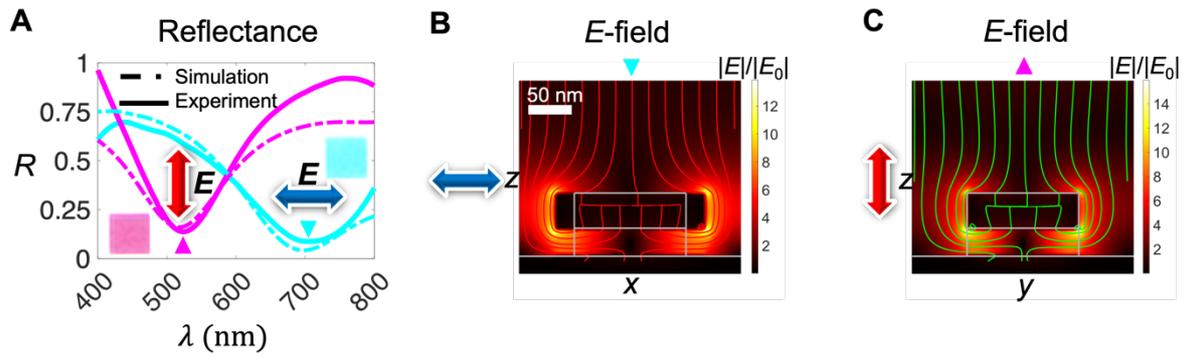

**Figure S1.** (**A**) Reflectance spectra of a color patch (inset) under *y* and *x*-polarized light. With *P* = 250 nm, $L_{xx}$ = 170 nm, $L_{yy}$ = 126 nm, $L_{xy}$ = 80 nm, $L_{yx}$ = 40 nm. (**B**) Electric field distribution overlaid with Poynting vector for the structure in the *x*-*z* plane under *x*-polarized light. (**C**) Electric field distribution overlaid with Poynting vector for the structure in the *y*-*z* plane under *y*-polarized light.

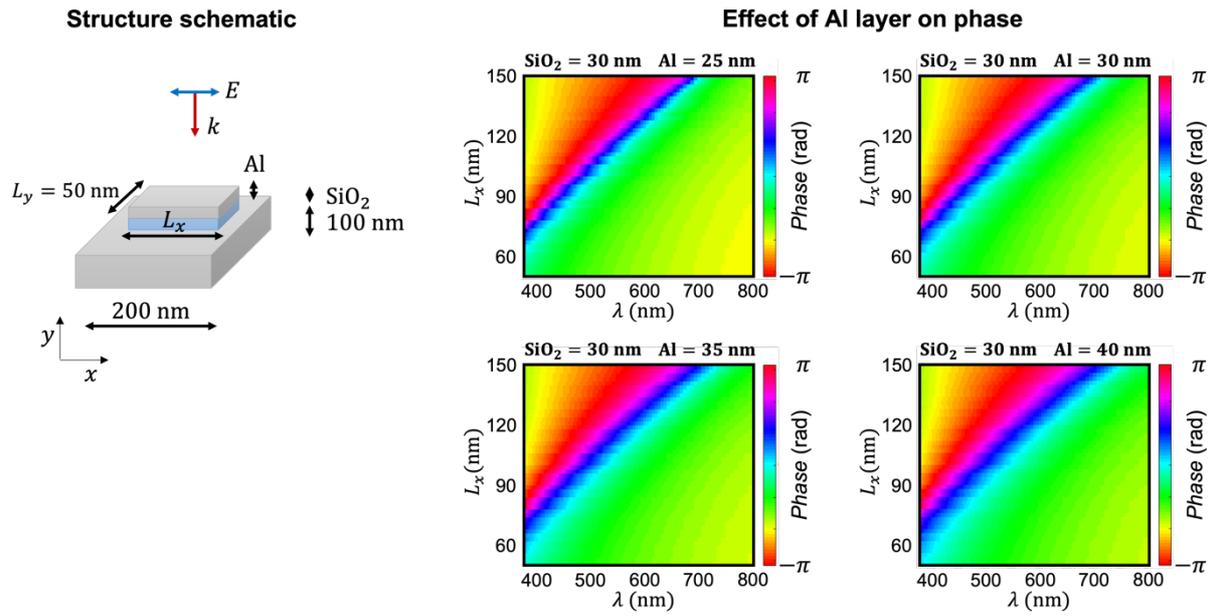

**Figure S2.** Effect of top Al layer on reflected phase.

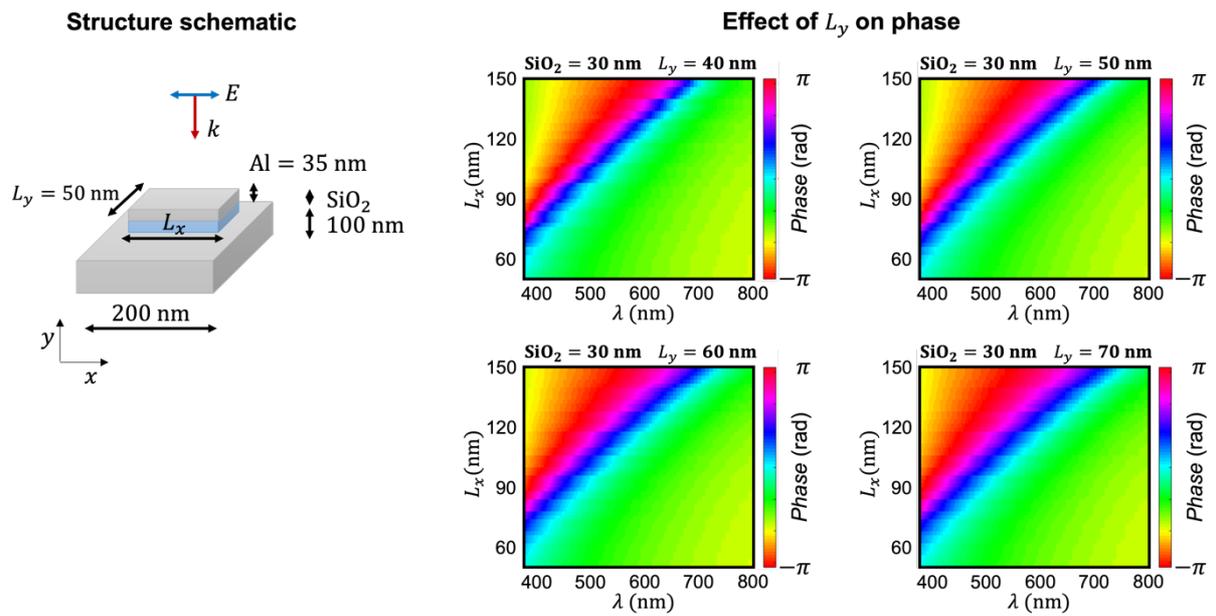

**Figure S3.** Effect of $L_y$ on reflected phase.

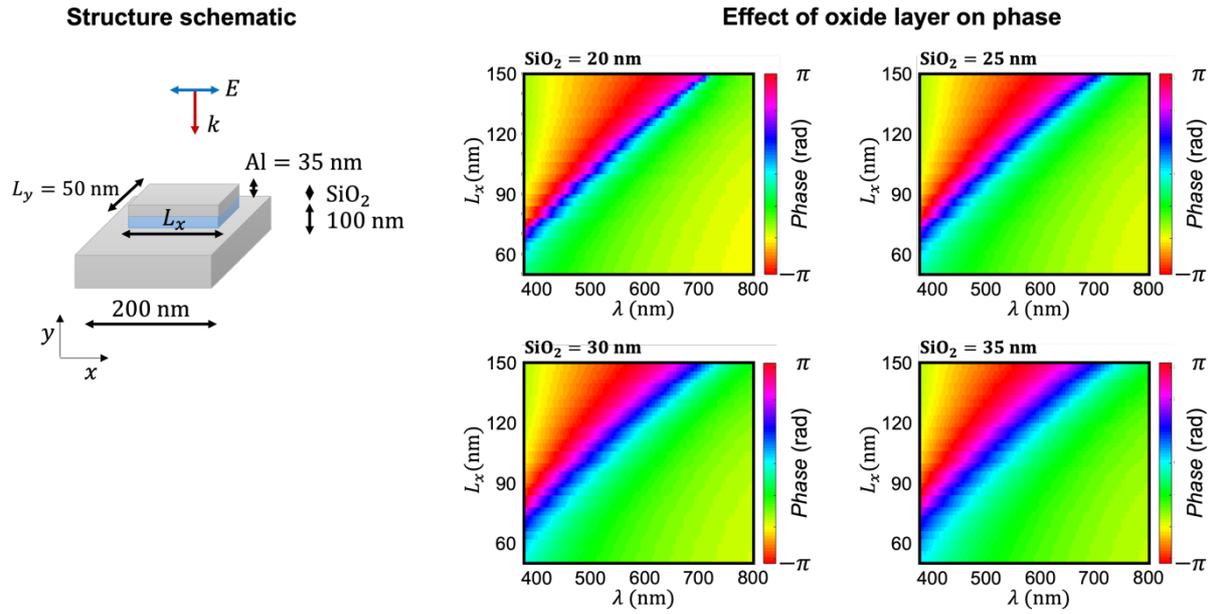

**Figure S4.** Effect of top SiO$_2$ layer on reflected phase.

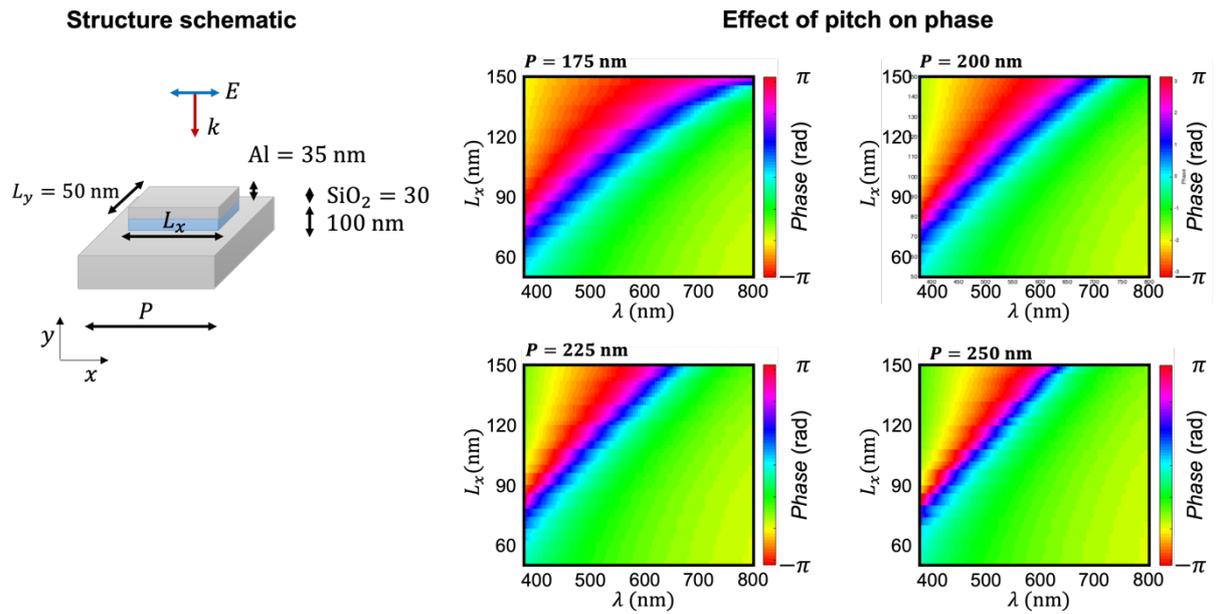

**Figure S5.** Effect of pitch $P$ on reflected phase.

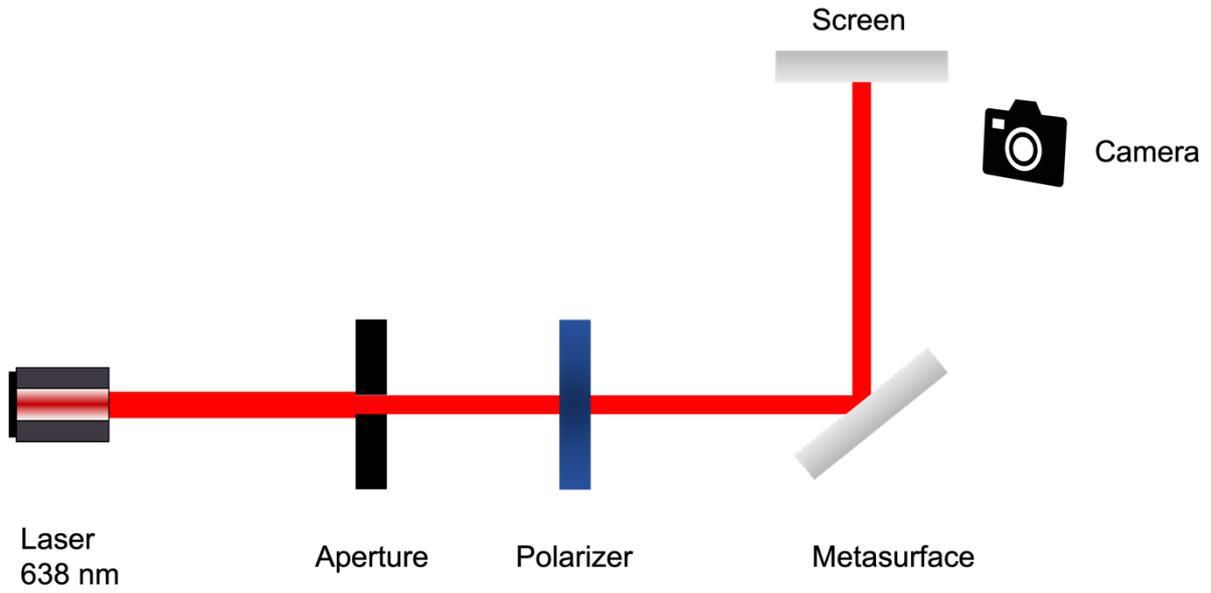

**Figure S6.** Experimental optical setup for capturing the hologram.

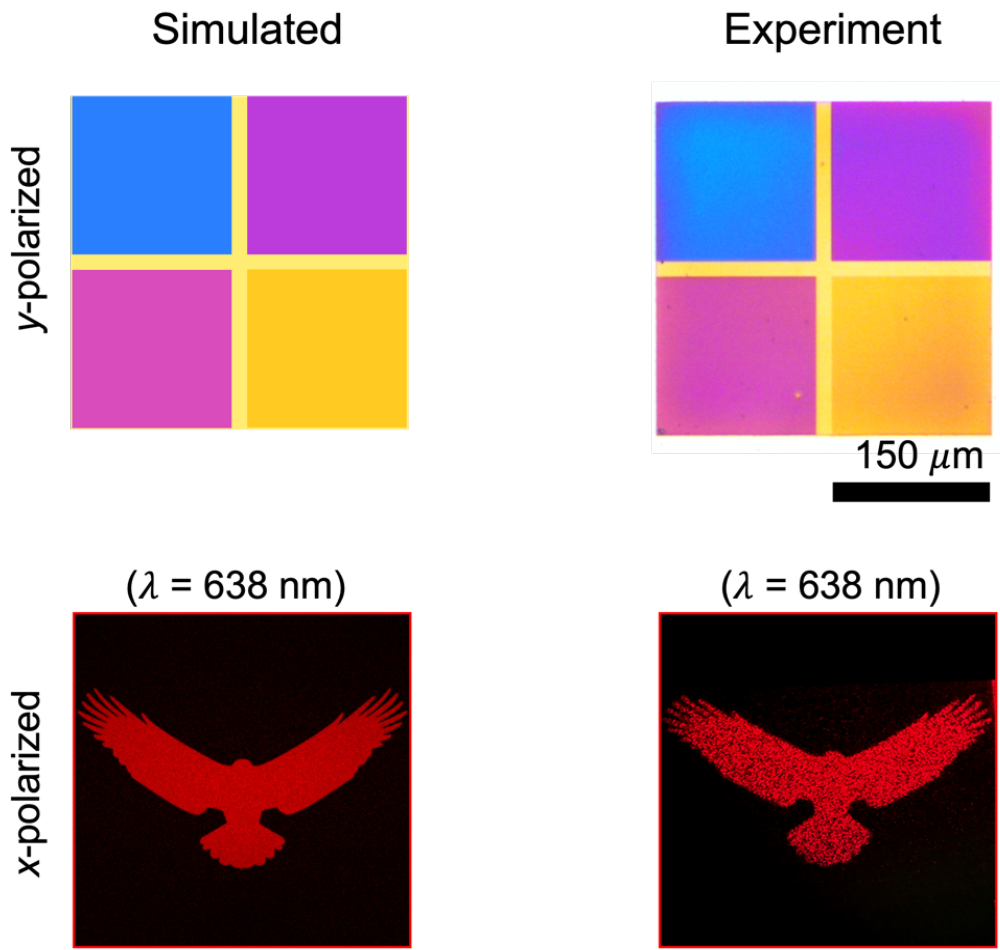

**Figure S7.** Simulated *vs.* experimental color print and hologram.

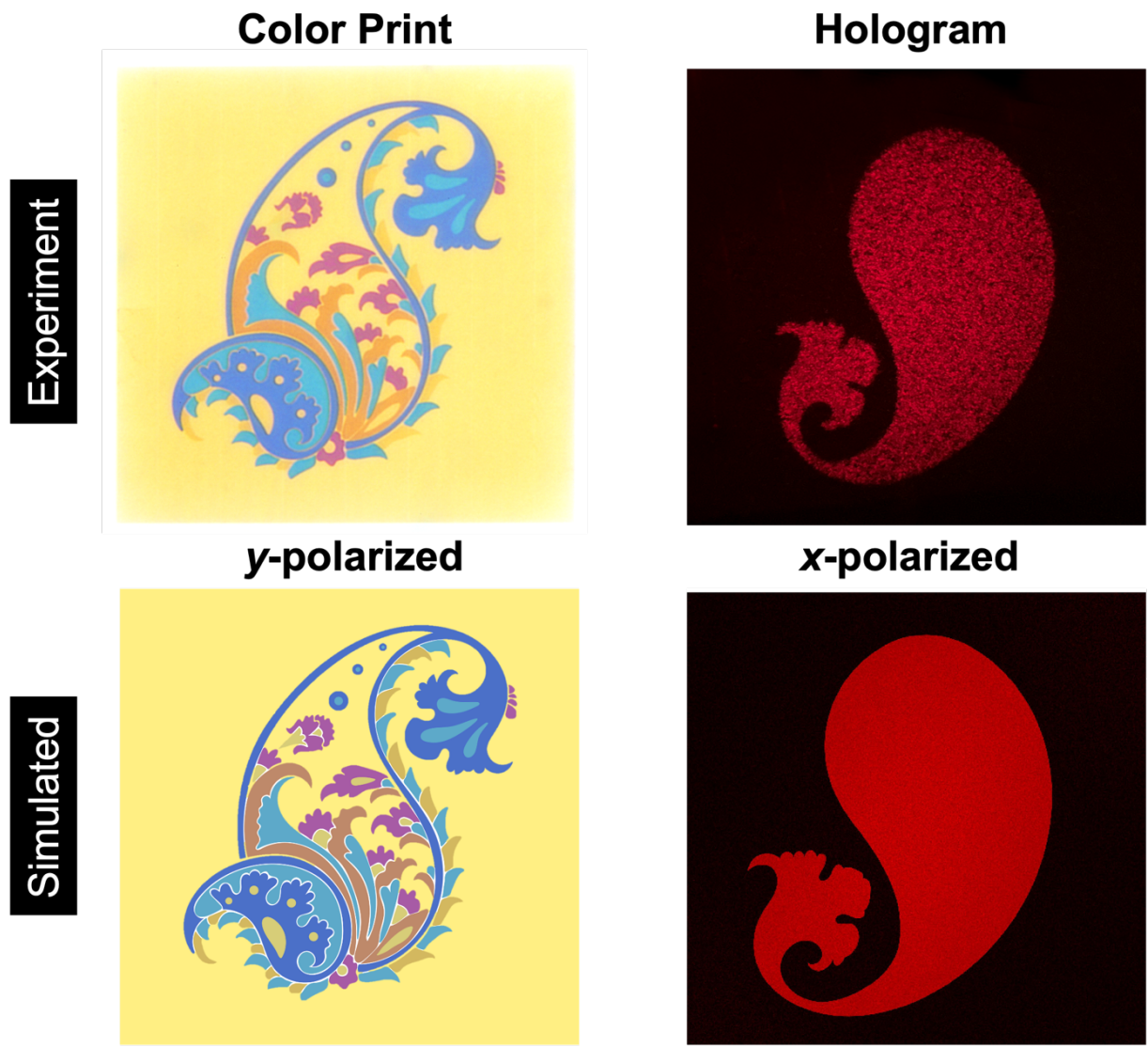

**Figure S8.** Simulated *vs*. experimental color print and hologram.

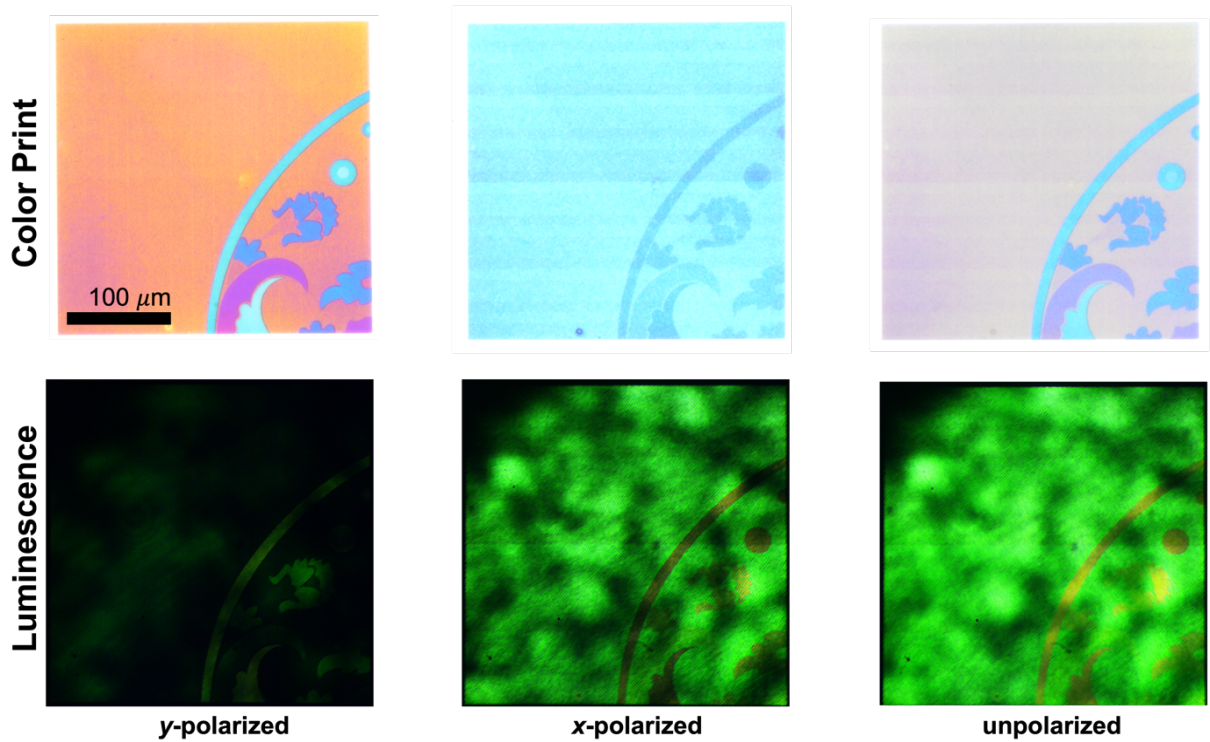

**Figure S9.** Color print and luminescence images under various polarizations.

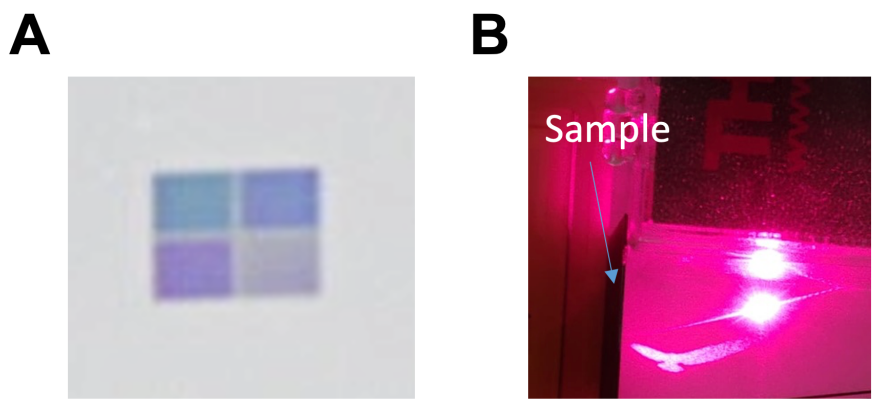

**Figure S10.** (**A**) Metasurface image captured by an iPhone 12 Pro Max augmented with a 25X macro lens (**B**) the hologram generated by the same sample using a red laser pointer.